\documentclass[reprint,amsmath,amssymb,aps,]{revtex4-2}

\usepackage{graphicx}
\usepackage{dcolumn}
\usepackage{bm}

\begin{document}

\preprint{APS/123-QED}

\title{Effect of Earth-Moon's gravity on TianQin's range acceleration noise. II. \\ Impact of orbit selection }

\author{Chengjian Luo}
\author{Xuefeng Zhang}
\email{zhangxf38@sysu.edu.cn}
\affiliation{
MOE Key Laboratory of TianQin Mission, TianQin Research Center for Gravitational Physics \& School of Physics and Astronomy, Frontiers Science Center for TianQin, Gravitational Wave Research Center of CNSA, Sun Yat-sen University (Zhuhai Campus), Zhuhai 519082, China.}

\date{\today}

\begin{abstract}

The paper is a sequel to our previous work (Zhang et al. Phys. Rev. D 103, 062001 (2021)). For proposed geocentric space-based gravitational wave detectors such as TianQin, gLISA, and GADFLI, the gravity-field disturbances, i.e., the so called ``orbital noise'', from the Earth-Moon system on the sensitive intersatellite laser interferometric measurements should be carefully evaluated and taken into account in the concept studies. Based on TianQin, we investigate how the effect, in terms of frequency spectra, varies with different choices of orbital orientations and radii through single-variable studies, and present the corresponding roll-off frequencies that may set the lower bounds of the targeted detection bands. The results, including the special cases of geostationary orbits (gLISA/GADFLI) and repeat orbits, can provide a useful input to orbit and constellation design for future geocentric missions. 

\end{abstract}

\date{\today}%
\maketitle

\section{Introduction}

In ground-based gravitational wave (GW) detectors such as LIGO \cite{Abramovici1992}, Virgo \cite{Caron1997}, and KAGRA \cite{Somiya2012}, Newtonian (or gravity-gradient) noise due to nearby terrestrial gravity fluctuations constitutes a prominent obstacle to sensitivity improvement below $\sim 10$ Hz \cite{Harms2019}. This difficulty on the ground, in addition to seismic noise and the need of long arms (typically $\sim 10^6$ km), has motivated space-based detection in order to access sub-Hertz GW signals \cite{Hild2012}. 

The space-based GW detection missions can be roughly categorized into heliocentric concepts, featured by LISA \cite{Folkner1997}, and geocentric concepts, featured by TianQin \cite{Luo2016}. For geocentric missions, though the issue of gravity-field disturbances is expected to be less severe, it may still require careful assessment due to the relative proximity of the satellites to the Earth and the Moon. The goal is to ascertain that there can be a proper separation in the frequencies of gravity-field disturbances and the targeted sensitive frequency bands near mHz, which is prerequisite to a successful space-based GW detection. 
Unlike ground-based detectors, armlengths of space-based detectors are generally not fixed primarily due to gravitational perturbations to orbital motion. Varying armlengths also induces Doppler shifts ($\sim 5$ MHz) in the beat notes of the inter-satellite heterodyne laser interferometric measurements. The frequency shifts may generate phase ramp-up, onto which GW-induced phase fluctuations are superimposed. The former, containing gravity-field information, should be removed or separated with caution in the pre-data processing stage on the ground. Thereby an accurate simulation of the orbital Doppler shifts would be needed to verify the phase measurement accuracy and time-delay interferometry \cite{Tinto2020}, at least for TianQin. 

The nominal orbit design for TianQin and other geocentric missions should take into account and trade off between a variety of factors \cite{Zhang2021b} such as science objectives, space environment, science payloads, etc. For example, in \cite{Tan2020}, the impact of orbit selection on the constellation stability of TianQin has been studied, and the permitted ranges of the orbital elements that can meet the payload requirements were identified. In a similar manner, one should also investigate the impact of orbit selection on gravity-field disturbances so as to facilitate trade studies in the orbit design and help seek potential remedies if needed. 

In the previous paper \cite{Zhang2021a}, we have shown that, with the $1\times 10^5$ km orbital radius and a vertical orbital plane relative to the ecliptic for TianQin, the Earth-Moon's gravity-field disturbances are dominating in lower frequencies, and roll off rapidly at $1\times 10^{-4}$ Hz. The effect does not enter the detection frequency band of $10^{-4} - 1$ Hz where the intersatellite residual acceleration noise requirement of $\sqrt{2}\times 10^{-15}$ m/s$^2$/Hz$^{1/2}$ is imposed, and hence presents no showstopper to the mission. For comparison, the gravity mapping mission GRACE-FO has a much higher roll-off frequency at $4\times 10^{-2}$ Hz \cite{Abich2019}, owing to its low altitude of $\sim 500$ km above the ground, at which the science instrument can 
be sensitive to the Earth's spherical harmonic gravity field to $>150$th degree \cite{Kornfeld2019}. Thereby, a separation of $1\times 10^5$ km from the Earth is considered large enough for TianQin to push the gravity-field disturbances out of the detection band. 

Nevertheless, these two isolated cases does not tell us much about other possible options in selecting orbital orientations and radii. It begs the questions on how the gravity-field effect changes with different orbit selections, and how one chooses a safe distance from gravity-field noise sources, particularly, the Earth. In this work, we intend to resolve these problems by evaluating the effect with respect to various inclinations, longitudes of ascending nodes, and orbital radii. The results may provide a useful reference to general orbit and constellation design for geocentric missions, such as GEOGRAWI/gLISA \cite{Tinto2015}, GADFLI \cite{McWilliams2011}, B-DECIGO \cite{Kawamura2018}, etc. 

This is our sixth paper of the concept study series on TianQin’s orbit and constellation \cite{Ye2019,Tan2020,Zhang2021a,Ye2021,Zhou2021}. For progress on other space environmental effects on TianQin, one may refer to \cite{Chen2021,Lu2021,Su2021,Lu2020,Su2020}. This article is structured as follows. Section \ref{sec:setup} describes the observable and force models used in the simulations. The results are shown in Section \ref{sec:result}, which also include geostationary orbits, and repeat orbits and resonant orbits are deferred to the Appendices. Finally, we reach the conclusions in Section \ref{sec:conclusion}. 

\section{Simulation Setup}\label{sec:setup}

Following \cite{Zhang2021a}, we adopt the range acceleration $\ddot\rho$ as the observable, which is simply the second time derivative of the range $\rho$ between two satellites. The criteria is to compare the amplitude spectral densities (ASD) of $\ddot\rho$ with the residual acceleration noise requirements between two free-floating test masses, which is generally in the order of $10^{-15}$ m/s$^2$/Hz$^{1/2}$ at mHz. To overcome the insufficiency of double precision, we again use the TQPOP (TianQin Quadruple Precision Orbit Propagator) program \cite{Zhang2021a} to propagate pure-gravity orbits in quadruple precision and with a constant step size of 50 seconds for 90 days. Here the step size is a balanced choice made among the factors of frequency band coverage, numerical accuracy, and computational run-time \cite{Zhang2021a}. Like TianQin, all constellations are assumed to be equilateral triangles barring small deviations, and the three satellites moving in the same nominal circular orbits with 120$^\circ$ phase separation (see, e.g., Table \ref{tab:orbit}). The spectral results are not sensitive to the specific year chosen. Without loss of generality, the initial epoch assumes 06 Jun. 2004, 00:00:00 UTC as in \cite{Zhang2021a}. 

\begin{table}[htb]
\caption{\label{tab:orbit} The nominal orbital elements of the TianQin constellation in the J2000-based Earth-centered ecliptic (EarthMJ2000Ec) coordinate system . Here $a$ denotes the semimajor axis, $e$ the eccentricity, $i$ the inclination, $\Omega$ the longitude of ascending node, $\omega$ the argument of periapsis, and $\Delta\nu$ the difference in the true anomaly.}
\begin{ruledtabular}
\begin{tabular}{ccccccc}
  & $a$ & $e$ & $i$ & $\Omega$ & $\omega$ & $\Delta\nu$ \\ 
\hline
SC1,2,3 & $1\times 10^5$ km & $0^\circ$ & $94.7^\circ$ & $210.4^\circ$ & $0^\circ$ & $120^\circ$ \\
\end{tabular}
\end{ruledtabular}
\end{table}

The detailed force models implemented in TQPOP are summarized in Table \ref{tab:models}. Further description of them can be found in \cite{Zhang2021a}. In particular, the program offers multiple options in modeling solar system ephemeris, Earth's precession and nutation, Earth's static gravity field, etc. The flexibility helps cross-checking model errors, which have been shown not to alter the spectral results for TianQin \cite{Zhang2021a}. In the following simulations, the more recent options will be used (except for the ocean tides). A new addition to TQPOP lately includes two ocean tide models FES2014b \cite{FES2014} and EOT11a \cite{Savcenko2012}. Based on the results of the test runs, all three models generate spectra consistent with one another in low frequencies. However, since FES2014b contains more high-frequency tidal components (e.g. M6, M8), we will use FES2014b in the following simulation. 

\begin{table}[htb]
\caption{\label{tab:models}
The list of force models implemented in TQPOP. If multiple options are given, the first one is used in the simulations. }
\begin{ruledtabular}
\begin{tabular}{lc}
Models                 & Specifications \\
\hline
Solar system ephemeris & JPL DE430 \cite{Folkner2014} \\
                       & JPL DE405 \cite{Standish1998} \\
\hline
Earth's precession \& nutation & IAU 2006/2000A \cite{Petit2010} \\
                               & IAU 1976/1980 \cite{McCarthy1996} \\
Earth's polar motion           & EOP 14 C04 \cite{EOP} \\
Earth's static gravity field   & EGM2008 ($n=12$) \cite{Pavlis2012} \\
                               & EGM96 ($n=12$)  \cite{Lemoine1998} \\
Solid Earth tides              & IERS (2010) \cite{Petit2010} \\
Ocean tides                    & FES2004 ($n=10$) \cite{Lyard2006} \\
                               & FES2014b ($n=10$) \cite{FES2014} \\
                               & EOT11a  ($n=10$) \cite{Savcenko2012} \\
Solid Earth pole tide          & IERS (2010) \cite{Petit2010} \\
Ocean pole tide                & Desai (2003) \cite{Petit2010} \\
Atmospheric tides              & Biancale \& Bode (2003) \cite{Biancale2006} \\ 
\hline
Moon's libration               & JPL DE430 \cite{Folkner2014} \\
                               & JPL DE405 \cite{Standish1998} \\
Moon's static gravity field    & GL0660B ($n=7$) \cite{Konopliv2013} \\
                               & LP165P ($n=7$) \cite{Konopliv2001} \\
Sun's orientation              & IAU \cite{Archinal2011}, Table 1 \\
Sun's $J_2$                    & IAU \cite{Archinal2011}, Table 1 \\
\hline
relativistic effect            & post-Newtonian \cite{McCarthy1996} \\
\end{tabular}
\end{ruledtabular}
\end{table}


\section{Simulation Results}\label{sec:result}

We can assess the impact of orbit selection on the gravity-field effect from the changes of the amplitudes $A_{kl}^{\pm}$ and frequencies $f_{kl}^{\pm}$ of spectral peaks in the range acceleration ASD curve. The frequencies $f_{kl}^{\pm}$ can be generally expressed as
\begin{eqnarray}\label{Exp:ASD}
f_{kl}^{\pm}=k\cdot f_{orb}\pm l\cdot f_{gra},
\end{eqnarray}
where $f_{orb}$ is the satellites' orbital frequency, and $f_{gra}$ is the intrinsic frequencies of gravity-field disturbances, and $k$, $l$ are integers.

The goal is to find how the gravity-field effect and the roll-off frequency vary with orbital elements. Here the roll-off frequency refers to where the range acceleration ASD curve intersects the acceleration noise requirement.

The study cases involve three orbital elements, i.e., the inclination $i$, the longitude of ascending node $\Omega$, and the semimajor axis $a$. Starting from the nominal orbit of TianQin given in Table \ref{tab:orbit}, we vary one single element at a time to make the trends more prominent. Besides, we discuss the particular case of geostationary orbits in Section \ref{sec:gLISA}. 


\subsection{Impact of orbital orientations}\label{sec:orientation}

Orbital inclinations $i$ and longitudes of ascending nodes $\Omega$ determine orbital orientations. For single-variable studies, we vary $i$ and $\Omega$ separately in the EarthMJ2000Ec coordinate system.

However, due to the existence of obliquity of the ecliptic, the influence of the Earth's static gravity field and its variation with orbital inclinations cannot be fully demonstrated in the EarthMJ2000Ec coordinate system by numerical method. Fortunately, from the previous work \cite{Zhang2021a} (Sec. IVB), we have learned that the total gravity-field effect in terms of the range-acceleration ASD can be roughly divided into two primary parts. The high-frequency part (above $\sim 5\times 10^{-5}$ Hz) is dominated by the non-spherical gravity field of the Earth, and the low-frequency part (below $\sim 5\times 10^{-5}$ Hz) is dominated by the Sun and the Moon. Therefore, based on symmetry considerations, we discuss the impact of the Earth's static gravity field with orbital orientations varying in the geocentric equatorial (EarthMJ2000Eq) coordinate system to supplement the effects of all the gravity-field disturbances in the EarthMJ2000Ec coordinate system. Here we mention that there is a $5^\circ$ difference between the Moon's orbital plane and the ecliptic plane, but it does not contribute significantly to the results. 

\begin{table}[htb]
\caption{\label{tab:inc} The orbital elements with different inclinations given in the EarthMJ2000Ec coordinate system at the epoch 06 Jun. 2004, 00:00:00 UTC. }
\begin{ruledtabular}
\begin{tabular}{ccc}
$a$ & $i$ & $\Omega$ \\
\hline
$1\times 10^5$ km & 0$^\circ$, 23.5$^\circ$, 45$^\circ$, 90$^\circ$, 135$^\circ$, 180$^\circ$ & 210.4$^\circ$ \\
\end{tabular}
\end{ruledtabular}
\end{table}

\begin{table}[htb]
\caption{\label{tab:RAAN} The orbital elements with different longitudes of ascending node given in the EarthMJ2000Ec coordinate system at the epoch 06 Jun. 2004, 00:00:00 UTC. }
\begin{ruledtabular}
\begin{tabular}{ccc}
$a$ & $i$ & $\Omega$ \\
\hline
$1\times 10^5$ km & 94.7$^\circ$ & 30.4$^\circ$, 120.4$^\circ$, 210.4$^\circ$, 300.4$^\circ$ \\
\end{tabular}
\end{ruledtabular}
\end{table}

In simulation, setting the orbital radius to a fixed value of $1\times 10^5$ km, we vary the inclination from $i=0^\circ$ to $180^\circ$ by $\Delta i=45^\circ$ intervals (see Table \ref{tab:inc}), where $i=23.5^\circ$ corresponds to the orbit near the equatorial plane. Then we take four distinct values of longitudes of ascending nodes $\Omega$ from $30.4^\circ$ to $300.4^\circ$ by $\Delta\Omega = 90^\circ$ intervals (Table \ref{tab:RAAN}) in the EarthMJ2000Ec coordinate system. Besides, to supplement the above cases, we vary the inclination from $i_{eq}=0^\circ$ to $180^\circ$ by $\Delta i_{eq}=30^\circ$ intervals (see Table \ref{tab:inc_eq}) in the EarthMJ2000Eq coordinate system. 
Due to rotational symmetry, the variation of the Earth's static gravity-field effect with different $\Omega_{eq}$ is insignificant, which is also confirmed by our simulations. A few observations can be summarized as follows. 

\begin{figure}[htb]
\includegraphics[scale=0.5]{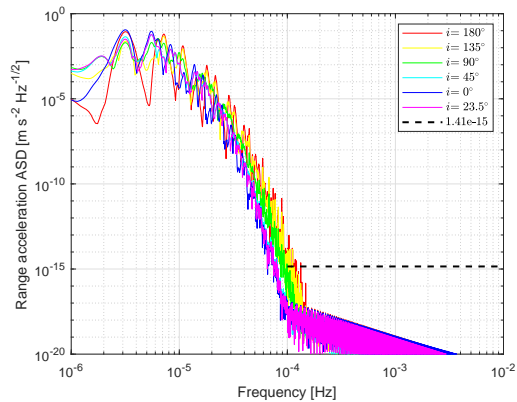}
\caption{\label{Fig:inc_ec} The range acceleration ASD between SC1 and SC2 calculated with the orbital parameters in Table \ref{tab:inc}, for 90 days with a constant step size of 50 seconds. }
\end{figure}

\begin{figure}[htb]
\includegraphics[scale=0.5]{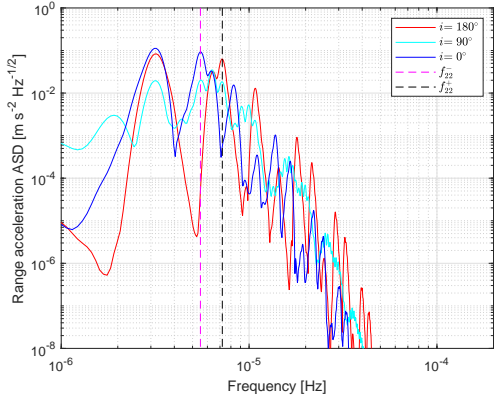}
\caption{\label{Fig:sun_moon_example} Examples of low-frequency peaks of Fig. \ref{Fig:inc_ec} and Eq. \ref{Exp:ASD}, where $f_{22}^{\pm}=2f_{orb}\pm2f_{moon}$. The plot shows that the amplitude $A_{22}^-$ falls and $A_{22}^+$ rises as the inclination increases in geocentric ecliptic coordinate system.}
\end{figure}

First, the low-frequency peaks of Fig. \ref{Fig:inc_ec} show that $A_{kl}^-$($l\neq0$) falls and $A_{kl}^+$($l\neq0$) rises as $i$ increases, which, for example, can be read from Fig. \ref{Fig:sun_moon_example}. Moreover, the peaks $k=1$, $l=0$ corresponding to the orbital frequencies coincide with different $i$'s in the plot. The difference in the amplitudes $A_{10}$ has been discussed in \cite{Tan2020}. 

\begin{table}[htb]
\caption{\label{tab:inc_eq} The orbital elements with different inclinations given in the EarthMJ2000Eq coordinate system at the epoch 06 Jun. 2004, 00:00:00 UTC, assuming circular orbits and $\Delta\nu =120^\circ$ for three satellites. Note that $i=94.7^\circ$, $\Omega=210.4^\circ$ (cf. Table \ref{tab:orbit}) correspond to $i_{eq}=74.5^\circ$, $\Omega_{eq}=211.6^\circ$. }
\begin{ruledtabular}
\begin{tabular}{ccc}
$a$ & $i_{eq}$ & $\Omega_{eq}$ \\
\hline
$1\times 10^5$ km & 0$^\circ$, 30$^\circ$, 60$^\circ$, 90$^\circ$, 120$^\circ$, 150$^\circ$, 180$^\circ$ & 211.6$^\circ$ \\
\end{tabular}
\end{ruledtabular}
\end{table}

\begin{figure}[htb]
\includegraphics[scale=0.5]{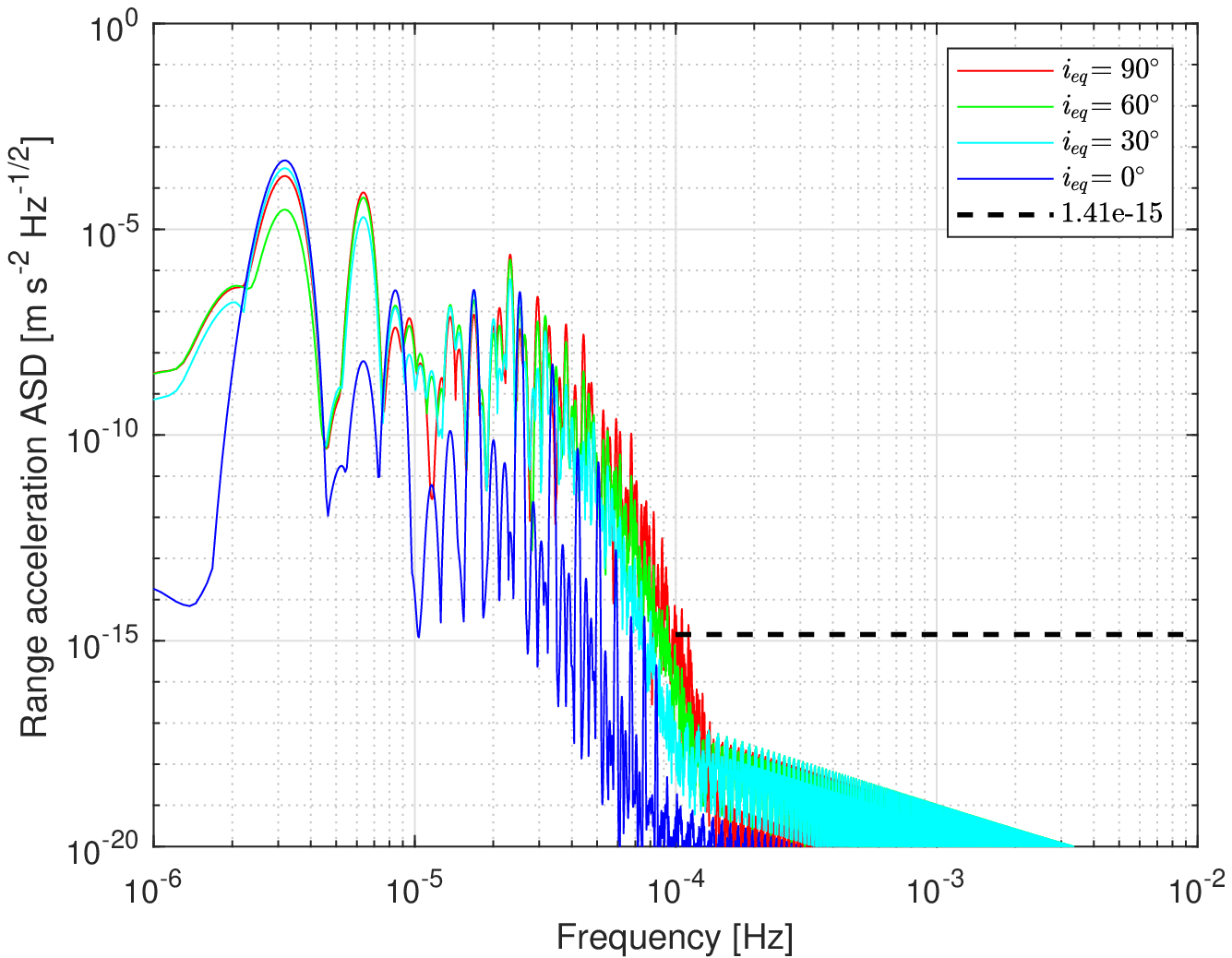}
\includegraphics[scale=0.5]{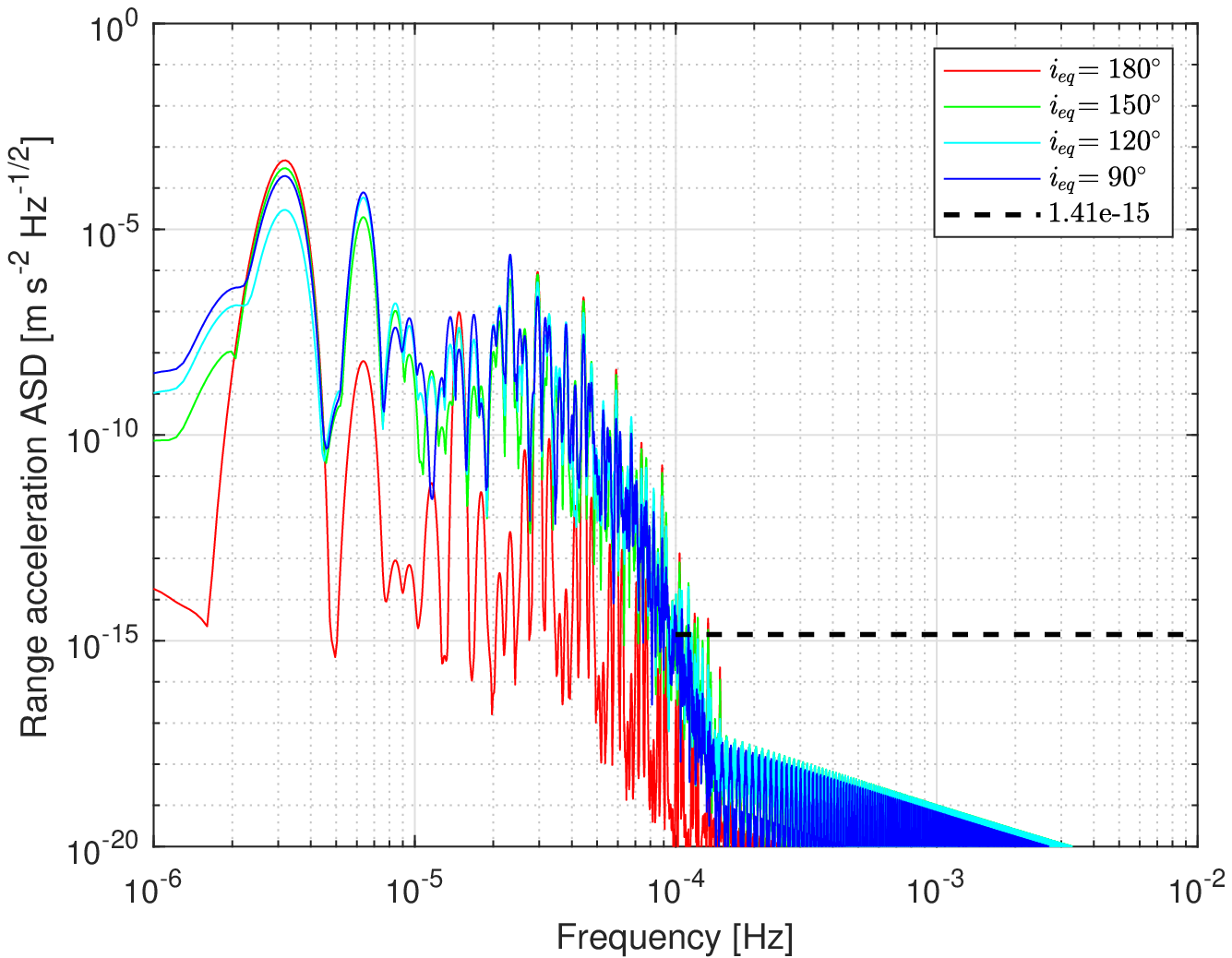}
\caption{\label{Fig:inc_eq_ns} The range acceleration ASD of the Earth's static gravity field between SC1 and SC2 calculated with the orbital parameters in Table \ref{tab:inc_eq}, for 90 days with a constant step size of 50 seconds.}
\end{figure}

Second, for high-frequency components, Fig. \ref{Fig:inc_ec} shows the ASD curves extend to higher frequencies monotonically as $i$ increases from $45^\circ$. Besides, Fig. \ref{Fig:inc_eq_ns} demonstrates that the ASD curves of the Earth's static gravity-field effect extend to higher frequencies as $i_{eq}$ increases. When $0^\circ\leq i\leq 45^\circ$ in the EarthMJ2000Ec coordinate system and $i_{eq}\geq 150^\circ$ in the EarthMJ2000Eq coordinate system, the roll-off frequency is insensitive to changes in inclinations, and hence it does not affect the range of the detection frequency band significantly. 

Third, when the satellites are running on the equatorial plane ($i_{eq}=0^\circ$ and $180^\circ$), Fig. \ref{Fig:inc_eq_ns} also shows the curves' trough values are obviously lower than other cases. This may be attributed to that the satellites experience less ``raggedness'' of the Earth's gravity field when orbiting in the equatorial plane (see also Fig. \ref{fig:gLISA_orbit}). 

\begin{figure}[htb]
 \includegraphics[scale=0.5]{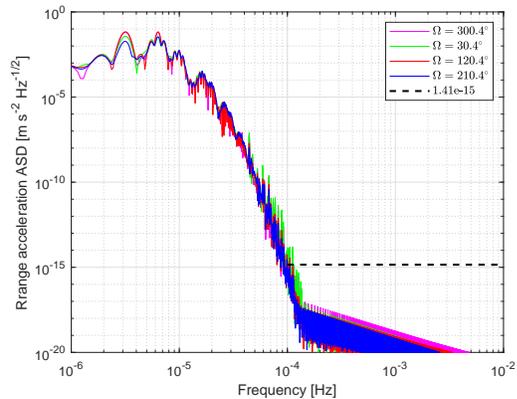}
 \caption{\label{Fig:raan} The range acceleration ASD between SC1 and SC2 calculated with the orbital parameters in Table \ref{tab:RAAN}, for 90 days with a constant step size of 50 seconds. }
\end{figure}

Fourth, compared with the case of orbital inclinations, the gravity-field effect is less sensitive to changes of the longitude of ascending node $\Omega$, as shown in Fig. \ref{Fig:raan}. In high frequencies, the differences come from that the Earth's static gravity field varies with $\Omega$ in the EarthMJ2000Ec coordinate system because of the obliquity of the ecliptic. In lower frequencies, the differences are partly due to the $5^\circ$ mismatch between the Moon's orbital plane and the ecliptic plane. The situation is somewhat similar to that of constellation stability \cite{Tan2020}, which is related to an approximate rotational symmetry about the ecliptic pole.

\subsection{Impact of orbital radii}\label{sec:radii}

Based on TianQin, we examine the orbital radius ranging from $4\times10^4$ km to $1.2\times10^5$ km by $\Delta a=2\times10^4$ km intervals. For special cases of repeat orbits and resonant orbits, see the appendices. The orbital elements are given in Table \ref{tab:radii}. 

\begin{table}[htb]
\caption{\label{tab:radii} The orbital elements with different orbital radii given in the EarthMJ2000Ec coordinate system at the epoch 06 Jun. 2004, 00:00:00 UTC. }
\begin{ruledtabular}
\begin{tabular}{ccc}
$a/10^4$ km & $i/^\circ$ & $\Omega/^\circ$ \\
\hline
4, 6, 8, 10, 12 & 94.7 &  210.4 \\ 
\end{tabular}
\end{ruledtabular}
\end{table}

\begin{figure}[htb]
\includegraphics[scale=0.5]{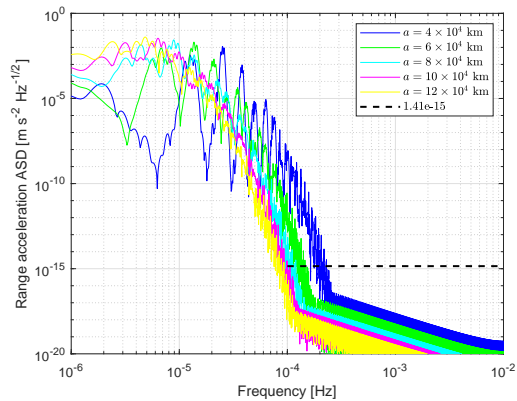}
\caption{\label{fig:radii} The range acceleration ASD with different orbital radii from Table \ref{tab:radii} and for 90 days with a constant step size of 50 seconds.}
\end{figure}

It should be noted that simulations with smaller orbital radii require higher degrees and orders in the Earth's gravity-field model. With estimated gravitational acceleration from different degrees, it is sufficient to set $n=16$ for $a=4\times 10^4$ km, $n=13$ for $a=6\times 10^4$ km, and $n=12$ for $a \geq 8\times 10^4$ km \cite{Zhang2021a}, so as to meet an accuracy below $10^{-16}$ m/s$^2$/Hz$^{1/2}$. The corresponding results of the range acceleration ASD are presented in Fig. \ref{fig:radii}.

\begin{figure}[htb]
\includegraphics[scale=0.5]{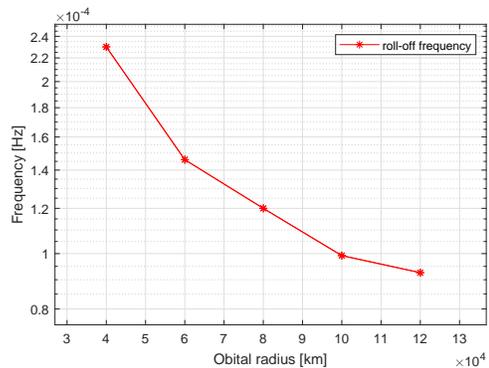}
\caption{\label{fig:relation_roll-off_and_radius} The relation between roll-off frequency and orbital radius of Fig. \ref{fig:radii}}
\end{figure}

First, with the orbital radius increasing, the ASD curves appear to shift to low frequencies overall. Moreover, one can read out the roll-off frequencies from Fig. \ref{fig:radii}, and the values are listed in Table \ref{Fit_coeffecient} and shown in Fig. \ref{fig:relation_roll-off_and_radius}. 

Second, one can fit the ASD curves of the range acceleration near the roll-off frequencies $f_r$ by
\begin{eqnarray}\label{fit}
\widetilde{RA}&=&\sqrt{2}\times10^{-15}\times\left( \frac{f_r}{f}\right)^{E_f} {\rm m/s^2/Hz^{1/2}},
\end{eqnarray}
where $f$ is the Fourier frequency, and $E_f$ the power index of the ratio. The fitted values of $E_f$ are given in Table \ref{Fit_coeffecient}. We mention that $E_f$ roughly changes with $a^2$ and $f_r$ with $1/a^2$. 

\begin{table}[htb]
\caption{\label{Fit_coeffecient}The numerical values of the roll-off frequency $f_r$ and the index $E_f$.}
\begin{ruledtabular}
\begin{tabular}{ccc}
   $a$/km   &   $f_r$/$10^{-4}$ Hz       &    $E_f$      \\\hline
   40000 &  2.30 &  21.0  \\
   60000 &  1.46 &  22.2  \\
   80000 &  1.20 &  23.8  \\
  100000 &  1.00 &  26.0  \\
  120000 &  0.93 &  28.6  \\
\end{tabular}
\end{ruledtabular}
\end{table}

\subsection{Geostationary orbits for gLISA/GADFLI}\label{sec:gLISA}

\begin{figure}[htb]
\includegraphics[scale=0.5]{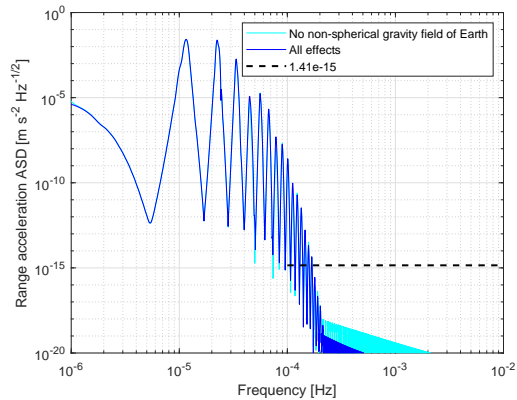}
\caption{\label{fig:gLISA_orbit} The range acceleration ASD between two satellites of gLISA/GADFLI, using the orbital parameters in Table \ref{gLISA orbit_parameters}, and calculated for 90 days with a constant step size of 50 seconds. }
\end{figure}

In this subsection, we examine geostationary orbits proposed by gLISA/GADFLI. The initial orbital elements are given in Table \ref{gLISA orbit_parameters}, which have been optimized for 90 days to stabilize the constellation. Here to focus on the gravity-field effect, we ignore station-keeping maneuvers that may be needed roughly every two weeks \cite{Tinto2015}. The corresponding range acceleration ASD is presented in Fig. \ref{fig:gLISA_orbit}, which incorporates the cases with and without the Earth's non-spherical gravity field in the simulation. 

Compared with Fig. \ref{fig:radii}, the plot shows a rather distinct feature in that the number of peaks is significantly reduced. This is due to the fact that by placing the satellites at the same location above the Earth's surface, the Earth's gravity experienced by the satellites becomes localized and nearly static. Therefore, the range acceleration ASD is instead dominated by the Moon and the Sun's gravity, with the high-frequency components from the Earth's non-spherical gravity field being suppressed. This is indicated by the overlapping of the blue and green curves in Fig. \ref{fig:gLISA_orbit}, where the latter is calculated without the Earth's non-spherical gravity field in orbit propagation. Moreover, the ASD curve intersects with the noise requirement $3\sqrt{2}\times10^{-15}$ m/s$^2$/Hz$^{1/2}$ at $1.69\times10^{-4}$ Hz, which is much lower than the proposed detection frequency band $10^{-2}-1$ Hz \cite{Tinto2015}.

\begin{table}[htb]
\caption{\label{gLISA orbit_parameters} The initial orbital elements of gLISA given in the EarthMJ2000Eq coordinate system at the epoch 06 Jun. 2004, 00:00:00 UTC. }
\begin{ruledtabular}
\begin{tabular}{ccccccc}
   Satellite &     $a$    &   $e$   &    $i$       &   $\Omega$    & $\omega$  & $\nu^{ini}$ \\ 
   \hline
   SC1      & 42158.65 km &         &              &               &           &  $210.0^\circ$ \\
   SC2      & 42164.90 km &   $0^\circ$    & $0^\circ$ &  $0^\circ$  & $0^\circ$ &  $330.8^\circ$ \\
   SC3      & 42171.00 km &         &              &               &           &  
   $\phantom{0}91.3^\circ$ \\ 
   \hline
\end{tabular}
\end{ruledtabular}
\end{table}

\section{Conclusion and Discussion}\label{sec:conclusion}

In this paper, we have simulated and analyzed the gravity-field effect of the Earth-Moon system on inter-satellite range acceleration of geocentric detectors in circular orbits with various orbital orientations and radii. It can help fill the blank in the concept studies of geocentric missions (e.g., \cite{NASA2012}) that might have been overlooked. As the previous work has shown \cite{Zhang2021a}, one common behavior is that the ``orbital noise'' dominates in the neighborhood of the orbital frequencies (e.g., $3\times 10^{-6}$ Hz for TianQin), and then falls off rapidly below residual acceleration noise requirements ($\sim 10^{-15}$ m/s$^2$/Hz$^{1/2}$) at specific higher frequencies. Besides, three main conclusions can be drawn here. 

First, for a fixed orbital radius $1\times 10^{5}$ km and the longitude of ascending node $210.4^\circ$, the roll-off frequency extends higher frequencies with the inclination increasing when the orbital inclination is greater than $45^\circ$ in the  EarthMJ2000Ec coordinate system. When the inclination is between $0^\circ$ and $45^\circ$ in the  EarthMJ2000Ec coordinate system or above $150^\circ$ in the  EarthMJ2000Eq coordinate system, the variation of roll-off frequency would not affect the choice of the detection frequency band significantly. 
For a given inclination, altering the longitude of ascending node has only a small impact.

Second, as the orbital radius increases, the gravity-field effect shifts to lower frequencies so that the detection frequency band can be extended accordingly (cf. Fig. \ref{fig:radii}). 

Third, the gravity-field effect for gLISA in geostationary orbits intersects with the acceleration noise requirement $3\sqrt{2}\times10^{-15}$ m/s$^2$/Hz$^{1/2}$ at $1.69\times10^{-4}$ Hz. It allows for sufficient clearance from the targeted detection band of $10^{-2}-1$ Hz \cite{Tinto2015}, and hence presents no showstopper to the mission. 

It should be pointed out that the results appear to favor small inclinations (see Section \ref{sec:orientation}). However, for TianQin, the inclination selection is largely driven by other more pressing technical demands. These include sciencecraft thermal control \cite{Zhang2018,Chen2021} and the need of lowering eclipse occurrence \cite{Ye2021}, etc., which instead favor a $90^\circ$ inclination. With a vertical orbital plane relative to the ecliptic, setting the radius equal to or greater than $1\times 10^5$ km is sufficient for TianQin to meet the requirement of extending the detection band to as low as $1\times 10^{-4}$ Hz. Meanwhile, the upper limit of the orbital radius is likely to be set by the constellation stability requirement \cite{Tan2020}. For future work, it would be desirable to develop an analytic model of the orbital dynamics to account for the range acceleration ASD at greater depths. 


\begin{acknowledgments}
The authors thank Yuzhou Fang, Lei Jiao, Bobing Ye, Jun Luo, and the anonymous referee for helpful discussions and comments. The work is supported by the National Key R\&D Program of China (No. 2020YFC2201202). X.Z. is supported by NSFC Grant No. 11805287.
\end{acknowledgments}

\appendix*

\section{A. Repeat orbits}

\begin{figure}[htb]
 \includegraphics[scale=0.5]{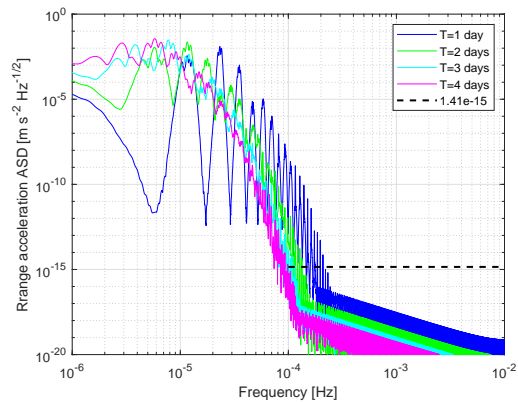}
 \caption{\label{fig:repeat_orbit}The range acceleration ASD of different repeat orbital radii between SC1 and SC2, using orbital parameters in Table \ref{tab:repeat_orbit_radii}, for 90 days with a constant step size of 50 seconds.}
\end{figure}

A repeat orbit has the property that the satellite covers the same ground track periodically. Such orbits may have certain engineering benefits. With the orbital plane facing J0806, we set the orbital periods from one day to four days, and give the initial orbital radii in Table \ref{tab:repeat_orbit_radii}. The corresponding results of the range acceleration ASD are presented in Fig. \ref{fig:repeat_orbit}. Notably, the ASD of the geosynchronous orbit ($T=1$ day) bears great resemblance to Fig. \ref{fig:gLISA_orbit}. 

\begin{table}[htb]
\caption{\label{tab:repeat_orbit_radii}The initial optimized orbital elements of different repeat orbits for the detectors in the EarthMJ2000Ec coordinate system at the epoch 06 Jun. 2004, 00:00:00 UTC.}
\begin{ruledtabular}
\begin{tabular}{cccc}
   $T$/day  &   SC1/km       &    SC2/km      &    SC3/km     \\\hline
  1  &  42164.14 &  42162.22 &  42163.60  \\
  2  &  66934.50 &  66936.30 &  66932.00  \\
  3  &  87714.00 &  87719.70 &  87710.00  \\
  4  & 106265.00 & 106278.50 &  106261.00  \\
\end{tabular}
\end{ruledtabular}
\end{table}

\section{B. Resonant orbits}
Although for TianQin, the work of \cite{Tan2020} has ruled out the orbits resonant with the Moon due to their inferior constellation stability performance. However, the results are still included for the sake of completeness. 

\begin{table}[htb]
\caption{\label{tab:Different_resonance_orbit_parameters}The initial orbital elements of different orbital resonance for the detectors in the EarthMJ2000Ec coordinate system at the epoch 06 Jun. 2004, 00:00:00 UTC.}
\begin{ruledtabular}
\begin{tabular}{cccc}
   $T_{moon}$/$T_{orb}$  &    SC1/km       &     SC2/km      &    SC3/km     \\\hline
  8  &  95796.617 &  95805.000 &  95791.000  \\
  7  & 104715.621 & 104728.000 & 104710.500  \\
  6  & 116049.338 & 116070.000 & 116040.000  \\
\end{tabular}
\end{ruledtabular}
\end{table}

\begin{figure}[htb]
 \includegraphics[scale=0.5]{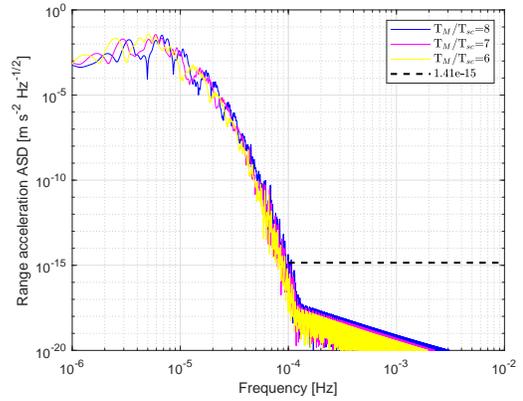}
 \caption{\label{fig:Different_resonance_orbit} The range acceleration ASD of different orbital resonance altitude between SC1 and SC2, using the orbital parameters in Table \ref{tab:Different_resonance_orbit_parameters}, for 90 days with a constant step size of 50 seconds.}
\end{figure}

We consider orbital period ratios $T_{moon}/T_{orb} = 6, 7, 8$. Table \ref{tab:Different_resonance_orbit_parameters} shows the corresponding orbital radii, and the other orbital elements are the same as Table \ref{tab:radii}. The results are presented in Fig. \ref{fig:Different_resonance_orbit}. Compared with non-resonant orbits, the peaks corresponding to the Earth-Moon coupling are enhanced. On the whole, the plot show a similar trend as Fig. \ref{fig:radii}. 

\bibliography{apscite}

\providecommand{\noopsort}[1]{}\providecommand{\singleletter}[1]{#1}%
\begin{thebibliography}{40}%
\makeatletter
\providecommand \@ifxundefined [1]{%
 \@ifx{#1\undefined}
}%
\providecommand \@ifnum [1]{%
 \ifnum #1\expandafter \@firstoftwo
 \else \expandafter \@secondoftwo
 \fi
}%
\providecommand \@ifx [1]{%
 \ifx #1\expandafter \@firstoftwo
 \else \expandafter \@secondoftwo
 \fi
}%
\providecommand \natexlab [1]{#1}%
\providecommand \enquote  [1]{``#1''}%
\providecommand \bibnamefont  [1]{#1}%
\providecommand \bibfnamefont [1]{#1}%
\providecommand \citenamefont [1]{#1}%
\providecommand \href@noop [0]{\@secondoftwo}%
\providecommand \href [0]{\begingroup \@sanitize@url \@href}%
\providecommand \@href[1]{\@@startlink{#1}\@@href}%
\providecommand \@@href[1]{\endgroup#1\@@endlink}%
\providecommand \@sanitize@url [0]{\catcode `\\12\catcode `\$12\catcode
  `\&12\catcode `\#12\catcode `\^12\catcode `\_12\catcode `\%12\relax}%
\providecommand \@@startlink[1]{}%
\providecommand \@@endlink[0]{}%
\providecommand \url  [0]{\begingroup\@sanitize@url \@url }%
\providecommand \@url [1]{\endgroup\@href {#1}{\urlprefix }}%
\providecommand \urlprefix  [0]{URL }%
\providecommand \Eprint [0]{\href }%
\providecommand \doibase [0]{https://doi.org/}%
\providecommand \selectlanguage [0]{\@gobble}%
\providecommand \bibinfo  [0]{\@secondoftwo}%
\providecommand \bibfield  [0]{\@secondoftwo}%
\providecommand \translation [1]{[#1]}%
\providecommand \BibitemOpen [0]{}%
\providecommand \bibitemStop [0]{}%
\providecommand \bibitemNoStop [0]{.\EOS\space}%
\providecommand \EOS [0]{\spacefactor3000\relax}%
\providecommand \BibitemShut  [1]{\csname bibitem#1\endcsname}%
\let\auto@bib@innerbib\@empty
\bibitem [{\citenamefont {Abramovici}\ \emph {et~al.}(1992)\citenamefont
  {Abramovici} \emph {et~al.}}]{Abramovici1992}%
  \BibitemOpen
  \bibfield  {author} {\bibinfo {author} {\bibfnamefont {A.}~\bibnamefont
  {Abramovici}} \emph {et~al.},\ }\bibfield  {title} {\bibinfo {title} {{LIGO}:
  the laser interferometer gravitational-wave observatory},\ }\href@noop {}
  {\bibfield  {journal} {\bibinfo  {journal} {Science}\ }\textbf {\bibinfo
  {volume} {256(5055)}},\ \bibinfo {pages} {325–333} (\bibinfo {year}
  {1992})}\BibitemShut {NoStop}%
\bibitem [{\citenamefont {Caron}\ \emph {et~al.}(1997)\citenamefont {Caron}
  \emph {et~al.}}]{Caron1997}%
  \BibitemOpen
  \bibfield  {author} {\bibinfo {author} {\bibfnamefont {B.}~\bibnamefont
  {Caron}} \emph {et~al.},\ }\bibfield  {title} {\bibinfo {title} {The {VIRGO}
  interferometer for gravitational wave detection},\ }\href@noop {} {\bibfield
  {journal} {\bibinfo  {journal} {Nucl. Phys. B, Proc. Suppl.}\ }\textbf
  {\bibinfo {volume} {54}},\ \bibinfo {pages} {167} (\bibinfo {year}
  {1997})}\BibitemShut {NoStop}%
\bibitem [{\citenamefont {Somiya}(2012)}]{Somiya2012}%
  \BibitemOpen
  \bibfield  {author} {\bibinfo {author} {\bibfnamefont {K.}~\bibnamefont
  {Somiya}},\ }\bibfield  {title} {\bibinfo {title} {Detector configuration of
  {KAGRA} - the japanese cryogenic gravitational-wave detector},\ }\href@noop
  {} {\bibfield  {journal} {\bibinfo  {journal} {Class. Quantum Gravity}\
  }\textbf {\bibinfo {volume} {29}},\ \bibinfo {pages} {124007} (\bibinfo
  {year} {2012})}\BibitemShut {NoStop}%
\bibitem [{\citenamefont {Harms}(2019)}]{Harms2019}%
  \BibitemOpen
  \bibfield  {author} {\bibinfo {author} {\bibfnamefont {J.}~\bibnamefont
  {Harms}},\ }\bibfield  {title} {\bibinfo {title} {Terrestrial gravity
  fluctuations},\ }\href@noop {} {\bibfield  {journal} {\bibinfo  {journal}
  {Living Rev. Relativ.}\ }\textbf {\bibinfo {volume} {22}},\ \bibinfo {pages}
  {6} (\bibinfo {year} {2019})}\BibitemShut {NoStop}%
\bibitem [{\citenamefont {Hild}(2012)}]{Hild2012}%
  \BibitemOpen
  \bibfield  {author} {\bibinfo {author} {\bibfnamefont {S.}~\bibnamefont
  {Hild}},\ }\bibfield  {title} {\bibinfo {title} {Beyond the second generation
  of laser interferometric gravitational wave observatories},\ }\href@noop {}
  {\bibfield  {journal} {\bibinfo  {journal} {Class. Quantum Grav.}\ }\textbf
  {\bibinfo {volume} {29}},\ \bibinfo {pages} {124006} (\bibinfo {year}
  {2012})}\BibitemShut {NoStop}%
\bibitem [{\citenamefont {Folkner}\ \emph {et~al.}(1997)\citenamefont
  {Folkner}, \citenamefont {Hechler}, \citenamefont {Sweetser}, \citenamefont
  {Vincent},\ and\ \citenamefont {Bender}}]{Folkner1997}%
  \BibitemOpen
  \bibfield  {author} {\bibinfo {author} {\bibfnamefont {W.~M.}\ \bibnamefont
  {Folkner}}, \bibinfo {author} {\bibfnamefont {F.}~\bibnamefont {Hechler}},
  \bibinfo {author} {\bibfnamefont {T.~H.}\ \bibnamefont {Sweetser}}, \bibinfo
  {author} {\bibfnamefont {M.~A.}\ \bibnamefont {Vincent}},\ and\ \bibinfo
  {author} {\bibfnamefont {P.~L.}\ \bibnamefont {Bender}},\ }\bibfield  {title}
  {\bibinfo {title} {{LISA} orbit selection and stability},\ }\href@noop {}
  {\bibfield  {journal} {\bibinfo  {journal} {Class. Quantum Grav.}\ }\textbf
  {\bibinfo {volume} {14}},\ \bibinfo {pages} {1405} (\bibinfo {year}
  {1997})}\BibitemShut {NoStop}%
\bibitem [{\citenamefont {Luo}\ \emph {et~al.}(2016)\citenamefont {Luo} \emph
  {et~al.}}]{Luo2016}%
  \BibitemOpen
  \bibfield  {author} {\bibinfo {author} {\bibfnamefont {J.}~\bibnamefont
  {Luo}} \emph {et~al.},\ }\bibfield  {title} {\bibinfo {title} {{TianQin}: a
  space-borne gravitational wave detector},\ }\href@noop {} {\bibfield
  {journal} {\bibinfo  {journal} {Class. Quantum Grav.}\ }\textbf {\bibinfo
  {volume} {33}},\ \bibinfo {pages} {035010} (\bibinfo {year}
  {2016})}\BibitemShut {NoStop}%
\bibitem [{\citenamefont {Tinto}\ and\ \citenamefont
  {Dhurandhar}(2021)}]{Tinto2020}%
  \BibitemOpen
  \bibfield  {author} {\bibinfo {author} {\bibfnamefont {M.}~\bibnamefont
  {Tinto}}\ and\ \bibinfo {author} {\bibfnamefont {S.~V.}\ \bibnamefont
  {Dhurandhar}},\ }\bibfield  {title} {\bibinfo {title} {Time-delay
  interferometry},\ }\href@noop {} {\bibfield  {journal} {\bibinfo  {journal}
  {Living Rev Relativ}\ }\textbf {\bibinfo {volume} {24}},\ \bibinfo {pages}
  {1} (\bibinfo {year} {2021})}\BibitemShut {NoStop}%
\bibitem [{\citenamefont {Zhang}\ \emph
  {et~al.}(2021{\natexlab{a}})\citenamefont {Zhang}, \citenamefont {Ye},
  \citenamefont {Tan}, \citenamefont {Yuan}, \citenamefont {Luo}, \citenamefont
  {Jiao}, \citenamefont {Gu}, \citenamefont {Ding},\ and\ \citenamefont
  {Mei}}]{Zhang2021b}%
  \BibitemOpen
  \bibfield  {author} {\bibinfo {author} {\bibfnamefont {X.}~\bibnamefont
  {Zhang}}, \bibinfo {author} {\bibfnamefont {B.}~\bibnamefont {Ye}}, \bibinfo
  {author} {\bibfnamefont {Z.}~\bibnamefont {Tan}}, \bibinfo {author}
  {\bibfnamefont {H.}~\bibnamefont {Yuan}}, \bibinfo {author} {\bibfnamefont
  {C.}~\bibnamefont {Luo}}, \bibinfo {author} {\bibfnamefont {L.}~\bibnamefont
  {Jiao}}, \bibinfo {author} {\bibfnamefont {D.}~\bibnamefont {Gu}}, \bibinfo
  {author} {\bibfnamefont {Y.}~\bibnamefont {Ding}},\ and\ \bibinfo {author}
  {\bibfnamefont {J.}~\bibnamefont {Mei}},\ }\bibfield  {title} {\bibinfo
  {title} {Orbit and constellation design for {TianQin}: progress review},\
  }\href@noop {} {\bibfield  {journal} {\bibinfo  {journal} {Acta Scientiarum
  Naturalium Universitatis Sunyatseni}\ }\textbf {\bibinfo {volume}
  {60(1-2)}},\ \bibinfo {pages} {123} (\bibinfo {year}
  {2021}{\natexlab{a}})}\BibitemShut {NoStop}%
\bibitem [{\citenamefont {Tan}\ \emph {et~al.}(2020)\citenamefont {Tan},
  \citenamefont {Ye},\ and\ \citenamefont {Zhang}}]{Tan2020}%
  \BibitemOpen
  \bibfield  {author} {\bibinfo {author} {\bibfnamefont {Z.}~\bibnamefont
  {Tan}}, \bibinfo {author} {\bibfnamefont {B.}~\bibnamefont {Ye}},\ and\
  \bibinfo {author} {\bibfnamefont {X.}~\bibnamefont {Zhang}},\ }\bibfield
  {title} {\bibinfo {title} {Impact of orbital orientations and radii on
  {TianQin} constellation stability},\ }\href@noop {} {\bibfield  {journal}
  {\bibinfo  {journal} {Int. J. Mod. Phys. D}\ }\textbf {\bibinfo {volume}
  {29}},\ \bibinfo {pages} {2050056} (\bibinfo {year} {2020})}\BibitemShut
  {NoStop}%
\bibitem [{\citenamefont {Zhang}\ \emph
  {et~al.}(2021{\natexlab{b}})\citenamefont {Zhang}, \citenamefont {Luo},
  \citenamefont {Jiao}, \citenamefont {Ye}, \citenamefont {Yuan}, \citenamefont
  {Cai}, \citenamefont {Gu}, \citenamefont {Mei},\ and\ \citenamefont
  {Luo}}]{Zhang2021a}%
  \BibitemOpen
  \bibfield  {author} {\bibinfo {author} {\bibfnamefont {X.}~\bibnamefont
  {Zhang}}, \bibinfo {author} {\bibfnamefont {C.}~\bibnamefont {Luo}}, \bibinfo
  {author} {\bibfnamefont {L.}~\bibnamefont {Jiao}}, \bibinfo {author}
  {\bibfnamefont {B.}~\bibnamefont {Ye}}, \bibinfo {author} {\bibfnamefont
  {H.}~\bibnamefont {Yuan}}, \bibinfo {author} {\bibfnamefont {L.}~\bibnamefont
  {Cai}}, \bibinfo {author} {\bibfnamefont {D.}~\bibnamefont {Gu}}, \bibinfo
  {author} {\bibfnamefont {J.}~\bibnamefont {Mei}},\ and\ \bibinfo {author}
  {\bibfnamefont {J.}~\bibnamefont {Luo}},\ }\bibfield  {title} {\bibinfo
  {title} {{Effect of Earth-Moon’s gravity on TianQin’s range acceleration
  noise}},\ }\href@noop {} {\bibfield  {journal} {\bibinfo  {journal} {Phys.
  Rev. D}\ }\textbf {\bibinfo {volume} {103}},\ \bibinfo {pages} {062001}
  (\bibinfo {year} {2021}{\natexlab{b}})}\BibitemShut {NoStop}%
\bibitem [{\citenamefont {Abich}\ \emph {et~al.}(2019)\citenamefont {Abich}
  \emph {et~al.}}]{Abich2019}%
  \BibitemOpen
  \bibfield  {author} {\bibinfo {author} {\bibfnamefont {K.}~\bibnamefont
  {Abich}} \emph {et~al.},\ }\bibfield  {title} {\bibinfo {title} {In-orbit
  performance of the {GRACE Follow-on} laser ranging interferometer},\
  }\href@noop {} {\bibfield  {journal} {\bibinfo  {journal} {Phys. Rev. Lett.}\
  }\textbf {\bibinfo {volume} {123}},\ \bibinfo {pages} {031101} (\bibinfo
  {year} {2019})}\BibitemShut {NoStop}%
\bibitem [{\citenamefont {Kornfeld}\ \emph {et~al.}(2019)\citenamefont
  {Kornfeld}, \citenamefont {Arnold}, \citenamefont {Gross}, \citenamefont
  {Dahya}, \citenamefont {Klipstein}, \citenamefont {Gath},\ and\ \citenamefont
  {Bettadpur}}]{Kornfeld2019}%
  \BibitemOpen
  \bibfield  {author} {\bibinfo {author} {\bibfnamefont {R.~P.}\ \bibnamefont
  {Kornfeld}}, \bibinfo {author} {\bibfnamefont {B.~W.}\ \bibnamefont
  {Arnold}}, \bibinfo {author} {\bibfnamefont {M.~A.}\ \bibnamefont {Gross}},
  \bibinfo {author} {\bibfnamefont {N.~T.}\ \bibnamefont {Dahya}}, \bibinfo
  {author} {\bibfnamefont {W.~M.}\ \bibnamefont {Klipstein}}, \bibinfo {author}
  {\bibfnamefont {P.~F.}\ \bibnamefont {Gath}},\ and\ \bibinfo {author}
  {\bibfnamefont {S.}~\bibnamefont {Bettadpur}},\ }\bibfield  {title} {\bibinfo
  {title} {Grace-fo: The gravity recovery and climate experiment follow-on
  mission},\ }\href {https://doi.org/10.2514/1.A34326} {\bibfield  {journal}
  {\bibinfo  {journal} {Journal of Spacecraft and Rockets}\ }\textbf {\bibinfo
  {volume} {56}},\ \bibinfo {pages} {931} (\bibinfo {year} {2019})},\ \Eprint
  {https://arxiv.org/abs/https://doi.org/10.2514/1.A34326}
  {https://doi.org/10.2514/1.A34326} \BibitemShut {NoStop}%
\bibitem [{\citenamefont {Tinto}\ \emph {et~al.}(2015)\citenamefont {Tinto},
  \citenamefont {DeBra}, \citenamefont {Buchman},\ and\ \citenamefont
  {Tilley}}]{Tinto2015}%
  \BibitemOpen
  \bibfield  {author} {\bibinfo {author} {\bibfnamefont {M.}~\bibnamefont
  {Tinto}}, \bibinfo {author} {\bibfnamefont {D.}~\bibnamefont {DeBra}},
  \bibinfo {author} {\bibfnamefont {S.}~\bibnamefont {Buchman}},\ and\ \bibinfo
  {author} {\bibfnamefont {S.}~\bibnamefont {Tilley}},\ }\bibfield  {title}
  {\bibinfo {title} {{gLISA}: geosynchronous laser interferometer space antenna
  concepts with off-the-shelf satellites},\ }\href@noop {} {\bibfield
  {journal} {\bibinfo  {journal} {Rev. Sci. Instrum.}\ }\textbf {\bibinfo
  {volume} {86}},\ \bibinfo {pages} {014501} (\bibinfo {year}
  {2015})}\BibitemShut {NoStop}%
\bibitem [{\citenamefont {McWilliams}(2011)}]{McWilliams2011}%
  \BibitemOpen
  \bibfield  {author} {\bibinfo {author} {\bibfnamefont {S.~T.}\ \bibnamefont
  {McWilliams}},\ }\href@noop {} {\bibinfo {title} {{Geostationary Antenna for
  Disturbance-Free Laser Interferometry (GADFLI)}}},\ \bibinfo {howpublished}
  {arXiv:1111.3708 [astro-ph.IM]} (\bibinfo {year} {2011})\BibitemShut
  {NoStop}%
\bibitem [{\citenamefont {Kawamura}\ \emph {et~al.}(2018)\citenamefont
  {Kawamura} \emph {et~al.}}]{Kawamura2018}%
  \BibitemOpen
  \bibfield  {author} {\bibinfo {author} {\bibfnamefont {S.}~\bibnamefont
  {Kawamura}} \emph {et~al.},\ }\bibfield  {title} {\bibinfo {title} {Space
  gravitational-wave antennas {DECIGO} and {B-DECIGO}},\ }\href@noop {}
  {\bibfield  {journal} {\bibinfo  {journal} {Int. J. Mod. Phys. D}\ }\textbf
  {\bibinfo {volume} {27}},\ \bibinfo {pages} {1845001} (\bibinfo {year}
  {2018})}\BibitemShut {NoStop}%
\bibitem [{\citenamefont {Ye}\ \emph {et~al.}(2019)\citenamefont {Ye},
  \citenamefont {Zhang}, \citenamefont {Zhou}, \citenamefont {Wang},
  \citenamefont {Yuan}, \citenamefont {Gu}, \citenamefont {Ding}, \citenamefont
  {Zhang}, \citenamefont {Mei},\ and\ \citenamefont {Luo}}]{Ye2019}%
  \BibitemOpen
  \bibfield  {author} {\bibinfo {author} {\bibfnamefont {B.}~\bibnamefont
  {Ye}}, \bibinfo {author} {\bibfnamefont {X.}~\bibnamefont {Zhang}}, \bibinfo
  {author} {\bibfnamefont {M.}~\bibnamefont {Zhou}}, \bibinfo {author}
  {\bibfnamefont {Y.}~\bibnamefont {Wang}}, \bibinfo {author} {\bibfnamefont
  {H.}~\bibnamefont {Yuan}}, \bibinfo {author} {\bibfnamefont {D.}~\bibnamefont
  {Gu}}, \bibinfo {author} {\bibfnamefont {Y.}~\bibnamefont {Ding}}, \bibinfo
  {author} {\bibfnamefont {J.}~\bibnamefont {Zhang}}, \bibinfo {author}
  {\bibfnamefont {J.}~\bibnamefont {Mei}},\ and\ \bibinfo {author}
  {\bibfnamefont {J.}~\bibnamefont {Luo}},\ }\bibfield  {title} {\bibinfo
  {title} {Optimizing orbits for {TianQin}},\ }\href@noop {} {\bibfield
  {journal} {\bibinfo  {journal} {Int. J. Mod. Phys. D}\ }\textbf {\bibinfo
  {volume} {28}},\ \bibinfo {pages} {1950121} (\bibinfo {year}
  {2019})}\BibitemShut {NoStop}%
\bibitem [{\citenamefont {Ye}\ \emph {et~al.}(2021)\citenamefont {Ye},
  \citenamefont {Zhang}, \citenamefont {Ding},\ and\ \citenamefont
  {Meng}}]{Ye2021}%
  \BibitemOpen
  \bibfield  {author} {\bibinfo {author} {\bibfnamefont {B.}~\bibnamefont
  {Ye}}, \bibinfo {author} {\bibfnamefont {X.}~\bibnamefont {Zhang}}, \bibinfo
  {author} {\bibfnamefont {Y.}~\bibnamefont {Ding}},\ and\ \bibinfo {author}
  {\bibfnamefont {Y.}~\bibnamefont {Meng}},\ }\bibfield  {title} {\bibinfo
  {title} {Eclipse avoidance in {TianQin} orbit selection},\ }\href@noop {}
  {\bibfield  {journal} {\bibinfo  {journal} {Phys. Rev. D}\ }\textbf {\bibinfo
  {volume} {103}},\ \bibinfo {pages} {042007} (\bibinfo {year}
  {2021})}\BibitemShut {NoStop}%
\bibitem [{\citenamefont {Zhou}\ \emph {et~al.}(2021)\citenamefont {Zhou},
  \citenamefont {Hu}, \citenamefont {Ye}, \citenamefont {Hu}, \citenamefont
  {Zhu}, \citenamefont {Zhang}, \citenamefont {Su}, ,\ and\ \citenamefont
  {Wang}}]{Zhou2021}%
  \BibitemOpen
  \bibfield  {author} {\bibinfo {author} {\bibfnamefont {M.}~\bibnamefont
  {Zhou}}, \bibinfo {author} {\bibfnamefont {X.}~\bibnamefont {Hu}}, \bibinfo
  {author} {\bibfnamefont {B.}~\bibnamefont {Ye}}, \bibinfo {author}
  {\bibfnamefont {S.}~\bibnamefont {Hu}}, \bibinfo {author} {\bibfnamefont
  {D.}~\bibnamefont {Zhu}}, \bibinfo {author} {\bibfnamefont {X.}~\bibnamefont
  {Zhang}}, \bibinfo {author} {\bibfnamefont {W.}~\bibnamefont {Su}}, ,\ and\
  \bibinfo {author} {\bibfnamefont {Y.}~\bibnamefont {Wang}},\ }\bibfield
  {title} {\bibinfo {title} {{Orbital effects on time delay interferometry for
  TianQin}},\ }\href@noop {} {\bibfield  {journal} {\bibinfo  {journal} {Phys.
  Rev. D}\ }\textbf {\bibinfo {volume} {103}},\ \bibinfo {pages} {103026}
  (\bibinfo {year} {2021})}\BibitemShut {NoStop}%
\bibitem [{\citenamefont {Chen}\ \emph {et~al.}(2021)\citenamefont {Chen},
  \citenamefont {Ling}, \citenamefont {Zhang}, \citenamefont {Zhao},
  \citenamefont {Li},\ and\ \citenamefont {Ding}}]{Chen2021}%
  \BibitemOpen
  \bibfield  {author} {\bibinfo {author} {\bibfnamefont {H.}~\bibnamefont
  {Chen}}, \bibinfo {author} {\bibfnamefont {C.}~\bibnamefont {Ling}}, \bibinfo
  {author} {\bibfnamefont {X.}~\bibnamefont {Zhang}}, \bibinfo {author}
  {\bibfnamefont {X.}~\bibnamefont {Zhao}}, \bibinfo {author} {\bibfnamefont
  {M.}~\bibnamefont {Li}},\ and\ \bibinfo {author} {\bibfnamefont
  {Y.}~\bibnamefont {Ding}},\ }\bibfield  {title} {\bibinfo {title} {{Thermal
  environment analysis for TianQin}},\ }\href@noop {} {\bibfield  {journal}
  {\bibinfo  {journal} {Class. Quantum Grav.}\ }\textbf {\bibinfo {volume}
  {38}},\ \bibinfo {pages} {155015} (\bibinfo {year} {2021})}\BibitemShut
  {NoStop}%
\bibitem [{\citenamefont {Lu}\ \emph {et~al.}(2021)\citenamefont {Lu},
  \citenamefont {Su}, \citenamefont {Zhang}, \citenamefont {He}, \citenamefont
  {Duan}, \citenamefont {Jiang},\ and\ \citenamefont {Yeh}}]{Lu2021}%
  \BibitemOpen
  \bibfield  {author} {\bibinfo {author} {\bibfnamefont {L.-F.}\ \bibnamefont
  {Lu}}, \bibinfo {author} {\bibfnamefont {W.}~\bibnamefont {Su}}, \bibinfo
  {author} {\bibfnamefont {X.}~\bibnamefont {Zhang}}, \bibinfo {author}
  {\bibfnamefont {Z.-G.}\ \bibnamefont {He}}, \bibinfo {author} {\bibfnamefont
  {H.-Z.}\ \bibnamefont {Duan}}, \bibinfo {author} {\bibfnamefont {Y.-Z.}\
  \bibnamefont {Jiang}},\ and\ \bibinfo {author} {\bibfnamefont {H.-C.}\
  \bibnamefont {Yeh}},\ }\bibfield  {title} {\bibinfo {title} {{Effects of the
  Space Plasma Density Oscillation on the Interspacecraft Laser Ranging for
  TianQin Gravitational Wave Observatory}},\ }\href@noop {} {\bibfield
  {journal} {\bibinfo  {journal} {JGR: Space Physics}\ }\textbf {\bibinfo
  {volume} {126}},\ \bibinfo {pages} {e2020JA028579} (\bibinfo {year}
  {2021})}\BibitemShut {NoStop}%
\bibitem [{\citenamefont {Su}\ \emph {et~al.}(2021)\citenamefont {Su},
  \citenamefont {Wang}, \citenamefont {Zhou}, \citenamefont {Lu}, \citenamefont
  {Zhou}, \citenamefont {Li}, \citenamefont {Shi}, \citenamefont {Hu},
  \citenamefont {Zhou}, \citenamefont {Wang}, \citenamefont {Yeh},
  \citenamefont {Wang},\ and\ \citenamefont {Chen}}]{Su2021}%
  \BibitemOpen
  \bibfield  {author} {\bibinfo {author} {\bibfnamefont {W.}~\bibnamefont
  {Su}}, \bibinfo {author} {\bibfnamefont {Y.}~\bibnamefont {Wang}}, \bibinfo
  {author} {\bibfnamefont {C.}~\bibnamefont {Zhou}}, \bibinfo {author}
  {\bibfnamefont {L.}~\bibnamefont {Lu}}, \bibinfo {author} {\bibfnamefont
  {Z.-B.}\ \bibnamefont {Zhou}}, \bibinfo {author} {\bibfnamefont
  {T.}~\bibnamefont {Li}}, \bibinfo {author} {\bibfnamefont {T.}~\bibnamefont
  {Shi}}, \bibinfo {author} {\bibfnamefont {X.-C.}\ \bibnamefont {Hu}},
  \bibinfo {author} {\bibfnamefont {M.-Y.}\ \bibnamefont {Zhou}}, \bibinfo
  {author} {\bibfnamefont {M.}~\bibnamefont {Wang}}, \bibinfo {author}
  {\bibfnamefont {H.-C.}\ \bibnamefont {Yeh}}, \bibinfo {author} {\bibfnamefont
  {H.}~\bibnamefont {Wang}},\ and\ \bibinfo {author} {\bibfnamefont
  {P.}~\bibnamefont {Chen}},\ }\bibfield  {title} {\bibinfo {title} {Analyses
  of laser propagation noises for {TianQin} gravitational wave observatory
  based on the global magnetosphere {MHD} simulations},\ }\href@noop {}
  {\bibfield  {journal} {\bibinfo  {journal} {ApJ}\ }\textbf {\bibinfo {volume}
  {914}},\ \bibinfo {pages} {139} (\bibinfo {year} {2021})}\BibitemShut
  {NoStop}%
\bibitem [{\citenamefont {LU}\ \emph {et~al.}(2020)\citenamefont {LU},
  \citenamefont {LIU}, \citenamefont {DUAN}, \citenamefont {JIANG},\ and\
  \citenamefont {YEH}}]{Lu2020}%
  \BibitemOpen
  \bibfield  {author} {\bibinfo {author} {\bibfnamefont {L.}~\bibnamefont
  {LU}}, \bibinfo {author} {\bibfnamefont {Y.}~\bibnamefont {LIU}}, \bibinfo
  {author} {\bibfnamefont {H.}~\bibnamefont {DUAN}}, \bibinfo {author}
  {\bibfnamefont {Y.}~\bibnamefont {JIANG}},\ and\ \bibinfo {author}
  {\bibfnamefont {H.-C.}\ \bibnamefont {YEH}},\ }\bibfield  {title} {\bibinfo
  {title} {Numerical simulations of the wavefront distortion of
  inter-spacecraft laser beams caused by solar wind and magnetospheric
  plasmas},\ }\href@noop {} {\bibfield  {journal} {\bibinfo  {journal} {Plasma
  Sci. Technol.}\ }\textbf {\bibinfo {volume} {22}},\ \bibinfo {pages} {115301}
  (\bibinfo {year} {2020})}\BibitemShut {NoStop}%
\bibitem [{\citenamefont {Su}\ \emph {et~al.}(2020)\citenamefont {Su},
  \citenamefont {Wang}, \citenamefont {Zhou}, \citenamefont {Bai},
  \citenamefont {Guo}, \citenamefont {Zhou}, \citenamefont {Lee}, \citenamefont
  {Wang}, \citenamefont {Zhou}, \citenamefont {Shi}, \citenamefont {Yin},\ and\
  \citenamefont {Zhang}}]{Su2020}%
  \BibitemOpen
  \bibfield  {author} {\bibinfo {author} {\bibfnamefont {W.}~\bibnamefont
  {Su}}, \bibinfo {author} {\bibfnamefont {Y.}~\bibnamefont {Wang}}, \bibinfo
  {author} {\bibfnamefont {Z.-B.}\ \bibnamefont {Zhou}}, \bibinfo {author}
  {\bibfnamefont {Y.-Z.}\ \bibnamefont {Bai}}, \bibinfo {author} {\bibfnamefont
  {Y.}~\bibnamefont {Guo}}, \bibinfo {author} {\bibfnamefont {C.}~\bibnamefont
  {Zhou}}, \bibinfo {author} {\bibfnamefont {T.}~\bibnamefont {Lee}}, \bibinfo
  {author} {\bibfnamefont {M.}~\bibnamefont {Wang}}, \bibinfo {author}
  {\bibfnamefont {M.-Y.}\ \bibnamefont {Zhou}}, \bibinfo {author}
  {\bibfnamefont {T.}~\bibnamefont {Shi}}, \bibinfo {author} {\bibfnamefont
  {H.}~\bibnamefont {Yin}},\ and\ \bibinfo {author} {\bibfnamefont {B.-T.}\
  \bibnamefont {Zhang}},\ }\bibfield  {title} {\bibinfo {title} {{Analyses of
  residual accelerations for TianQin based on the global MHD simulation}},\
  }\href@noop {} {\bibfield  {journal} {\bibinfo  {journal} {Class. Quant.
  Grav.}\ }\textbf {\bibinfo {volume} {37}},\ \bibinfo {pages} {185017}
  (\bibinfo {year} {2020})}\BibitemShut {NoStop}%
\bibitem [{FES()}]{FES2014}%
  \BibitemOpen
  \href@noop {} {}\Eprint
  {https://arxiv.org/abs/https://grace.obs-mip.fr/dealiasing\_and\_tides/ocean-tides/}
  {https://grace.obs-mip.fr/dealiasing\_and\_tides/ocean-tides/} \BibitemShut
  {NoStop}%
\bibitem [{\citenamefont {Savcenko}\ and\ \citenamefont
  {Bosch}(2012)}]{Savcenko2012}%
  \BibitemOpen
  \bibfield  {author} {\bibinfo {author} {\bibfnamefont {R.}~\bibnamefont
  {Savcenko}}\ and\ \bibinfo {author} {\bibfnamefont {W.}~\bibnamefont
  {Bosch}},\ }\href@noop {} {\emph {\bibinfo {title} {EOT11a - empirical ocean
  tide model from multi-mission satellite altimetry}}},\ \bibinfo {type} {Tech.
  Rep.}\ (\bibinfo  {institution} {Deutsches Geodätisches Forschungsinstitut
  (DGFI)},\ \bibinfo {year} {2012})\BibitemShut {NoStop}%
\bibitem [{\citenamefont {Folkner}\ \emph {et~al.}(2014)\citenamefont
  {Folkner}, \citenamefont {Williams}, \citenamefont {Boggs}, \citenamefont
  {Park},\ and\ \citenamefont {Kuchynka}}]{Folkner2014}%
  \BibitemOpen
  \bibfield  {author} {\bibinfo {author} {\bibfnamefont {W.~M.}\ \bibnamefont
  {Folkner}}, \bibinfo {author} {\bibfnamefont {J.~G.}\ \bibnamefont
  {Williams}}, \bibinfo {author} {\bibfnamefont {D.~H.}\ \bibnamefont {Boggs}},
  \bibinfo {author} {\bibfnamefont {R.~S.}\ \bibnamefont {Park}},\ and\
  \bibinfo {author} {\bibfnamefont {P.}~\bibnamefont {Kuchynka}},\ }\href@noop
  {} {\emph {\bibinfo {title} {The planetary and lunar ephemerides {DE430} and
  {DE431}}}},\ \bibinfo {type} {IPN Progress Report}\ \bibinfo {number}
  {42-196}\ (\bibinfo  {institution} {Jet Propulsion Laboratory},\ \bibinfo
  {year} {2014})\BibitemShut {NoStop}%
\bibitem [{\citenamefont {Standish}(1998)}]{Standish1998}%
  \BibitemOpen
  \bibfield  {author} {\bibinfo {author} {\bibfnamefont {E.}~\bibnamefont
  {Standish}},\ }\href@noop {} {\emph {\bibinfo {title} {JPL Planetary and
  Lunar Ephemerides, DE405/LE405}}},\ \bibinfo {type} {Tech. Rep.}\ \bibinfo
  {number} {312.F-98-048}\ (\bibinfo  {institution} {JPL IOM},\ \bibinfo {year}
  {1998})\BibitemShut {NoStop}%
\bibitem [{\citenamefont {Petit}\ and\ \citenamefont
  {Luzum}(2010)}]{Petit2010}%
  \BibitemOpen
  \bibfield  {author} {\bibinfo {author} {\bibfnamefont {G.}~\bibnamefont
  {Petit}}\ and\ \bibinfo {author} {\bibfnamefont {B.}~\bibnamefont {Luzum}},\
  }\href@noop {} {\emph {\bibinfo {title} {IERS Conventions (2010)}}},\
  \bibinfo {type} {Technical Report}\ \bibinfo {number} {36}\ (\bibinfo
  {institution} {BUREAU INTERNATIONAL DES POIDS ET MESURES SEVRES (FRANCE)},\
  \bibinfo {year} {2010})\BibitemShut {NoStop}%
\bibitem [{\citenamefont {McCarthy}(1996)}]{McCarthy1996}%
  \BibitemOpen
  \bibfield  {author} {\bibinfo {author} {\bibfnamefont {D.~D.}\ \bibnamefont
  {McCarthy}},\ }\href@noop {} {\emph {\bibinfo {title} {IERS Conventions
  (1996)}}},\ \bibinfo {type} {Technical Report}\ \bibinfo {number} {21}\
  (\bibinfo  {institution} {Central Bureau of IERS - Observatoire de Paris},\
  \bibinfo {year} {1996})\BibitemShut {NoStop}%
\bibitem [{EOP()}]{EOP}%
  \BibitemOpen
  \href@noop {} {}\Eprint
  {https://arxiv.org/abs/http://hpiers.obspm.fr/iers/eop/eopc04/}
  {http://hpiers.obspm.fr/iers/eop/eopc04/} \BibitemShut {NoStop}%
\bibitem [{\citenamefont {Pavlis}\ \emph {et~al.}(2012)\citenamefont {Pavlis},
  \citenamefont {Holmes}, \citenamefont {Kenyon},\ and\ \citenamefont
  {Factor}}]{Pavlis2012}%
  \BibitemOpen
  \bibfield  {author} {\bibinfo {author} {\bibfnamefont {N.~K.}\ \bibnamefont
  {Pavlis}}, \bibinfo {author} {\bibfnamefont {S.~A.}\ \bibnamefont {Holmes}},
  \bibinfo {author} {\bibfnamefont {S.~C.}\ \bibnamefont {Kenyon}},\ and\
  \bibinfo {author} {\bibfnamefont {J.~K.}\ \bibnamefont {Factor}},\ }\bibfield
   {title} {\bibinfo {title} {The development and evaluation of the {Earth
  Gravitational Model 2008} ({EGM2008})},\ }\href@noop {} {\bibfield  {journal}
  {\bibinfo  {journal} {J. Geophys. Res.}\ }\textbf {\bibinfo {volume} {117}},\
  \bibinfo {pages} {B04406} (\bibinfo {year} {2012})}\BibitemShut {NoStop}%
\bibitem [{\citenamefont {Lemoine}\ \emph {et~al.}(1998)\citenamefont
  {Lemoine}, \citenamefont {Kenyon}, \citenamefont {Factor}, \citenamefont
  {Trimmer}, \citenamefont {Pavlis}, \citenamefont {Chinn}, \citenamefont
  {Cox}, \citenamefont {Klosko}, \citenamefont {Luthcke}, \citenamefont
  {Torrence}, \citenamefont {Wang}, \citenamefont {Williamson}, \citenamefont
  {Pavlis}, \citenamefont {Rapp},\ and\ \citenamefont {Olson}}]{Lemoine1998}%
  \BibitemOpen
  \bibfield  {author} {\bibinfo {author} {\bibfnamefont {F.~G.}\ \bibnamefont
  {Lemoine}}, \bibinfo {author} {\bibfnamefont {S.~C.}\ \bibnamefont {Kenyon}},
  \bibinfo {author} {\bibfnamefont {J.~K.}\ \bibnamefont {Factor}}, \bibinfo
  {author} {\bibfnamefont {R.}~\bibnamefont {Trimmer}}, \bibinfo {author}
  {\bibfnamefont {N.~K.}\ \bibnamefont {Pavlis}}, \bibinfo {author}
  {\bibfnamefont {D.~S.}\ \bibnamefont {Chinn}}, \bibinfo {author}
  {\bibfnamefont {C.~M.}\ \bibnamefont {Cox}}, \bibinfo {author} {\bibfnamefont
  {S.~M.}\ \bibnamefont {Klosko}}, \bibinfo {author} {\bibfnamefont {S.~B.}\
  \bibnamefont {Luthcke}}, \bibinfo {author} {\bibfnamefont {M.~H.}\
  \bibnamefont {Torrence}}, \bibinfo {author} {\bibfnamefont {Y.~M.}\
  \bibnamefont {Wang}}, \bibinfo {author} {\bibfnamefont {R.~G.}\ \bibnamefont
  {Williamson}}, \bibinfo {author} {\bibfnamefont {E.~C.}\ \bibnamefont
  {Pavlis}}, \bibinfo {author} {\bibfnamefont {R.~H.}\ \bibnamefont {Rapp}},\
  and\ \bibinfo {author} {\bibfnamefont {T.~R.}\ \bibnamefont {Olson}},\
  }\href@noop {} {\emph {\bibinfo {title} {The Development of the Joint NASA
  GSFC and NIMA Geopotential Model EGM96}}},\ \bibinfo {type} {Tech. Rep.}\
  (\bibinfo  {institution} {NASA Goddard Space Flight Center},\ \bibinfo {year}
  {1998})\BibitemShut {NoStop}%
\bibitem [{\citenamefont {Lyard}\ \emph {et~al.}(2006)\citenamefont {Lyard},
  \citenamefont {Lefevre}, \citenamefont {Letellier},\ and\ \citenamefont
  {Francis}}]{Lyard2006}%
  \BibitemOpen
  \bibfield  {author} {\bibinfo {author} {\bibfnamefont {F.}~\bibnamefont
  {Lyard}}, \bibinfo {author} {\bibfnamefont {F.}~\bibnamefont {Lefevre}},
  \bibinfo {author} {\bibfnamefont {T.}~\bibnamefont {Letellier}},\ and\
  \bibinfo {author} {\bibfnamefont {O.}~\bibnamefont {Francis}},\ }\bibfield
  {title} {\bibinfo {title} {Modelling the global ocean tides: modern insights
  from {FES2004}},\ }\href@noop {} {\bibfield  {journal} {\bibinfo  {journal}
  {Ocean Dyn.}\ }\textbf {\bibinfo {volume} {56}},\ \bibinfo {pages} {394}
  (\bibinfo {year} {2006})}\BibitemShut {NoStop}%
\bibitem [{\citenamefont {Biancale}\ and\ \citenamefont
  {Bode}(2006)}]{Biancale2006}%
  \BibitemOpen
  \bibfield  {author} {\bibinfo {author} {\bibfnamefont {R.}~\bibnamefont
  {Biancale}}\ and\ \bibinfo {author} {\bibfnamefont {A.}~\bibnamefont
  {Bode}},\ }\href@noop {} {\emph {\bibinfo {title} {Mean annual and seasonal
  atmospheric tide models based on 3-hourly and 6-hourly ECMWF surface pressure
  data}}},\ \bibinfo {type} {STR}\ \bibinfo {number} {06/01}\ (\bibinfo
  {institution} {GeoForschungsZentrum Potsdam},\ \bibinfo {year}
  {2006})\BibitemShut {NoStop}%
\bibitem [{\citenamefont {Konopliv}\ \emph {et~al.}(2013)\citenamefont
  {Konopliv}, \citenamefont {Park}, \citenamefont {Yuan}, \citenamefont
  {Asmar}, \citenamefont {Watkins}, \citenamefont {Williams}, \citenamefont
  {Fahnestock}, \citenamefont {Kruizinga}, \citenamefont {Paik}, \citenamefont
  {Strekalov}, \citenamefont {Harvey}, \citenamefont {Smith},\ and\
  \citenamefont {Zuber}}]{Konopliv2013}%
  \BibitemOpen
  \bibfield  {author} {\bibinfo {author} {\bibfnamefont {A.~S.}\ \bibnamefont
  {Konopliv}}, \bibinfo {author} {\bibfnamefont {R.~S.}\ \bibnamefont {Park}},
  \bibinfo {author} {\bibfnamefont {D.-N.}\ \bibnamefont {Yuan}}, \bibinfo
  {author} {\bibfnamefont {S.~W.}\ \bibnamefont {Asmar}}, \bibinfo {author}
  {\bibfnamefont {M.~M.}\ \bibnamefont {Watkins}}, \bibinfo {author}
  {\bibfnamefont {J.~G.}\ \bibnamefont {Williams}}, \bibinfo {author}
  {\bibfnamefont {E.}~\bibnamefont {Fahnestock}}, \bibinfo {author}
  {\bibfnamefont {G.}~\bibnamefont {Kruizinga}}, \bibinfo {author}
  {\bibfnamefont {M.}~\bibnamefont {Paik}}, \bibinfo {author} {\bibfnamefont
  {D.}~\bibnamefont {Strekalov}}, \bibinfo {author} {\bibfnamefont
  {N.}~\bibnamefont {Harvey}}, \bibinfo {author} {\bibfnamefont {D.~E.}\
  \bibnamefont {Smith}},\ and\ \bibinfo {author} {\bibfnamefont {M.~T.}\
  \bibnamefont {Zuber}},\ }\bibfield  {title} {\bibinfo {title} {The {JPL}
  lunar gravity field to spherical harmonic degree 660 from the {GRAIL} primary
  mission},\ }\href@noop {} {\bibfield  {journal} {\bibinfo  {journal} {J.
  Geophys. Res. Planets}\ }\textbf {\bibinfo {volume} {118}},\ \bibinfo {pages}
  {1} (\bibinfo {year} {2013})}\BibitemShut {NoStop}%
\bibitem [{\citenamefont {Konopliv}\ \emph {et~al.}(2001)\citenamefont
  {Konopliv}, \citenamefont {Asmar}, \citenamefont {Carranza}, \citenamefont
  {Sjogren},\ and\ \citenamefont {Yuan}}]{Konopliv2001}%
  \BibitemOpen
  \bibfield  {author} {\bibinfo {author} {\bibfnamefont {A.}~\bibnamefont
  {Konopliv}}, \bibinfo {author} {\bibfnamefont {S.}~\bibnamefont {Asmar}},
  \bibinfo {author} {\bibfnamefont {E.}~\bibnamefont {Carranza}}, \bibinfo
  {author} {\bibfnamefont {W.}~\bibnamefont {Sjogren}},\ and\ \bibinfo {author}
  {\bibfnamefont {D.}~\bibnamefont {Yuan}},\ }\bibfield  {title} {\bibinfo
  {title} {Recent gravity models as a result of the lunar prospector mission},\
  }\href@noop {} {\bibfield  {journal} {\bibinfo  {journal} {Icarus}\ }\textbf
  {\bibinfo {volume} {150}},\ \bibinfo {pages} {1} (\bibinfo {year}
  {2001})}\BibitemShut {NoStop}%
\bibitem [{\citenamefont {Archinal}\ \emph {et~al.}(2011)\citenamefont
  {Archinal}, \citenamefont {A'Hearn}, \citenamefont {Bowell}, \citenamefont
  {Conrad}, \citenamefont {Consolmagno}, \citenamefont {Courtin}, \citenamefont
  {Fukushima}, \citenamefont {Hestroffer}, \citenamefont {Hilton},
  \citenamefont {Krasinsky}, \citenamefont {Neumann}, \citenamefont {Oberst},
  \citenamefont {Seidelmann}, \citenamefont {Stooke}, \citenamefont {Tholen},
  \citenamefont {Thomas},\ and\ \citenamefont {Williams}}]{Archinal2011}%
  \BibitemOpen
  \bibfield  {author} {\bibinfo {author} {\bibfnamefont {B.~A.}\ \bibnamefont
  {Archinal}}, \bibinfo {author} {\bibfnamefont {M.~F.}\ \bibnamefont
  {A'Hearn}}, \bibinfo {author} {\bibfnamefont {E.}~\bibnamefont {Bowell}},
  \bibinfo {author} {\bibfnamefont {A.}~\bibnamefont {Conrad}}, \bibinfo
  {author} {\bibfnamefont {G.~J.}\ \bibnamefont {Consolmagno}}, \bibinfo
  {author} {\bibfnamefont {R.}~\bibnamefont {Courtin}}, \bibinfo {author}
  {\bibfnamefont {T.}~\bibnamefont {Fukushima}}, \bibinfo {author}
  {\bibfnamefont {D.}~\bibnamefont {Hestroffer}}, \bibinfo {author}
  {\bibfnamefont {J.~L.}\ \bibnamefont {Hilton}}, \bibinfo {author}
  {\bibfnamefont {G.~A.}\ \bibnamefont {Krasinsky}}, \bibinfo {author}
  {\bibfnamefont {G.}~\bibnamefont {Neumann}}, \bibinfo {author} {\bibfnamefont
  {J.}~\bibnamefont {Oberst}}, \bibinfo {author} {\bibfnamefont {P.~K.}\
  \bibnamefont {Seidelmann}}, \bibinfo {author} {\bibfnamefont
  {P.}~\bibnamefont {Stooke}}, \bibinfo {author} {\bibfnamefont {D.~J.}\
  \bibnamefont {Tholen}}, \bibinfo {author} {\bibfnamefont {P.~C.}\
  \bibnamefont {Thomas}},\ and\ \bibinfo {author} {\bibfnamefont {I.~P.}\
  \bibnamefont {Williams}},\ }\bibfield  {title} {\bibinfo {title} {Report of
  the {IAU} working group on cartographic coordinates and rotational elements:
  2009},\ }\href@noop {} {\bibfield  {journal} {\bibinfo  {journal} {Celest.
  Mech. Dyn. Astr.}\ }\textbf {\bibinfo {volume} {109}},\ \bibinfo {pages}
  {101} (\bibinfo {year} {2011})}\BibitemShut {NoStop}%
\bibitem [{\citenamefont {Anderson}\ \emph {et~al.}(2012)\citenamefont
  {Anderson}, \citenamefont {Stebbins}, \citenamefont {Weiss},\ and\
  \citenamefont {Wright}}]{NASA2012}%
  \BibitemOpen
  \bibfield  {author} {\bibinfo {author} {\bibfnamefont {K.}~\bibnamefont
  {Anderson}}, \bibinfo {author} {\bibfnamefont {R.}~\bibnamefont {Stebbins}},
  \bibinfo {author} {\bibfnamefont {R.}~\bibnamefont {Weiss}},\ and\ \bibinfo
  {author} {\bibfnamefont {E.}~\bibnamefont {Wright}},\ }\href
  {https://www.cosmos.esa.int/documents/427239/442202/GW\_Study\_Rev3\_Aug2012-Final-1.pdf}
  {\emph {\bibinfo {title} {{NASA} {G}ravitational-wave mission concept study
  final report}}} (\bibinfo {year} {2012})\BibitemShut {NoStop}%
\bibitem [{\citenamefont {Zhang}\ \emph {et~al.}(2018)\citenamefont {Zhang},
  \citenamefont {Li},\ and\ \citenamefont {Mei}}]{Zhang2018}%
  \BibitemOpen
  \bibfield  {author} {\bibinfo {author} {\bibfnamefont {X.}~\bibnamefont
  {Zhang}}, \bibinfo {author} {\bibfnamefont {H.}~\bibnamefont {Li}},\ and\
  \bibinfo {author} {\bibfnamefont {J.}~\bibnamefont {Mei}},\ }\bibfield
  {title} {\bibinfo {title} {{Thermal stability estimation of TianQin
  satellites based on LISA-like thermal design concept}}} (\bibinfo {year}
  {2018}),\ \bibinfo {note} {{internal technical report} (in
  Chinese)}\BibitemShut {NoStop}%
\end{thebibliography}%

\end{document}